\documentclass[sigconf]{acmart}
\makeatletter
\def\@ACM@checkaffil{% Only warnings
    \if@ACM@instpresent\else
    \ClassWarningNoLine{\@classname}{No institution present for an affiliation}%
    \fi
    \if@ACM@citypresent\else
    \ClassWarningNoLine{\@classname}{No city present for an affiliation}%
    \fi
    \if@ACM@countrypresent\else
        \ClassWarningNoLine{\@classname}{No country present for an affiliation}%
    \fi
}
\makeatother

\renewcommand\footnotetextcopyrightpermission[1]{} % removes footnote with conference information in first column
\AtBeginDocument{%
  \providecommand\BibTeX{{%
    \normalfont B\kern-0.5em{\scshape i\kern-0.25em b}\kern-0.8em\TeX}}}

\usepackage{longtable}
\usepackage[utf8]{inputenc}
\usepackage{amsmath}
\usepackage{graphicx}
\usepackage[export]{adjustbox}
\usepackage{subcaption}
\usepackage{siunitx}
\usepackage{booktabs}
\usepackage{xcolor}
\usepackage[draft,inline,nomargin,index]{fixme}
\newtheorem{definition}{Definition}[section]

\begin{document}

\title{County-level Algorithmic Audit of Racial Bias in Twitter's Home Timeline}

\author{Luca Belli}
\orcid{0000-0002-2749-0586}
\affiliation{%
  \institution{Twitter Inc.}
}
\author{Kyra Yee}
\affiliation{%
  \institution{Twitter Inc.}
 }
 
 \author{Uthaipon Tantipongpipat}
\affiliation{%
  \institution{Twitter Inc.}
 }
 \author{Aaron Gonzales}
\affiliation{%
  \institution{Twitter Inc.}
 }
 \author{Kristian Lum}
\affiliation{%
  \institution{Twitter Inc.}
 }
 \author{Moritz Hardt
 }
 
\affiliation{%
  \institution{Max Planck Institute for Intelligent Systems, Tübingen}
 }
 \authornote{Work performed while MH was a paid consultant for Twitter.}
% Luca Belli, Kyra Yee, Uthaipon Tantipongpipat, Aaron Gonzales, Kristian Lum, Moritz Hardt

%\renewcommand{\shortauthors}{Belli, et al.}

\begin{abstract}
We report on the outcome of an audit of Twitter's Home Timeline ranking system. The goal of the audit was to determine if authors from some racial groups experience systematically higher impression counts for their Tweets than others.

A central obstacle for any such audit is that Twitter does not ordinarily collect or associate racial information with its users, thus prohibiting an analysis at the level of individual authors.  Working around this obstacle, we take US counties as our unit of analysis. We associate each user in the United States on the Twitter platform to a county based on available location data. The US Census Bureau provides information about the racial decomposition of the population in each county. 

The question we investigate then is if the racial decomposition of a county is associated with the visibility of Tweets originating from within the county. Focusing on two racial groups, the Black or African American population and the White population as defined by the US Census Bureau, we evaluate two statistical measures of bias.

Our investigation represents the first large-scale algorithmic audit into racial bias on the Twitter platform. Additionally, it illustrates the challenges of measuring racial bias in online platforms without having such information on the users. 
\end{abstract}




\maketitle
\pagestyle{plain}
\section{Introduction}

The content on Twitter’s Home Timeline is by default selected and ordered by personalization algorithms. This raises questions of algorithmic biases, such as: Does algorithmic personalization increase the visibility of some groups relative to others in a manner that aligns with and reinforces persisting inequality? In this paper, we audit Twitter's Home Timeline ranking algorithm for racial bias. The scope of our study is limited to the United States and focuses on two racial groups: the Black or African American population and the White population, as defined by the US Census Bureau.

Twitter does not generally collect or associate racial information with its users, which poses a methodological challenge to our audit. In fact at no time was any individual user's actual race or ethnicity known or used as part of the research. Addressing this challenge, we use coarse location information associated with a user that places the primary location of a user within a US county.
Counties vary in their racial demographics, raising the possibility that differences in a county’s visibility on Twitter systematically align with its racial demographics. Although there are challenges in using aggregate level data as proxies for individual level disparities \cite{freedman1999ecological,soobader2001using,geronimus1996validity}, county-level data has been used to study racial disparities in a variety of domains, such as law enforcement \cite{ross2015multi}, healthcare \cite{monnat2019using,gornick1996effects,leitner2016racial}, and education \cite{riddle2019racial}. There is also growing interest to applying county level data to understanding disparate impact in algorithmic applications \cite{johnson2017effect,chang2021targeted,kallus2022assessing}, where individual level demographic data is often difficult to procure \cite{andrus2021we}.

Choosing counties as our unit of analysis, we study two non-experimental ways of measuring bias. The first considers the percentage of \emph{amplified users} in each county, i.e., users whose Tweets have higher normalized impression counts than the US median user. We test whether the racial demographics of a county have an effect on the fraction of amplified users in the county. The second analysis splits counties based on whether the county has a greater than median population within a specific racial group. For the two groups of counties, the analysis then compares the distribution of amplified users.

Both analyses suggest that the size of the Black or African American population within a county is not significantly associated with higher or lower normalized impression counts. Both analyses show a small, but statistically significant, negative correlation between the size of the White population within a county and normalized impressions on the Home Timeline.

The contributions of our report are as follows:

\begin{itemize}
    \item We extend the growing body of work on algorithm audits to include a large-scale study of racial bias on the Twitter platform.
    \item We show one way to work around general unavailability of user-level racial information in the context of an algorithm audit, while detailing the limitations of this approach.
    \item Our report provides systematic quantitative information about racial bias on the Twitter platform.
    \item Along with this report, we release the county-level dataset used for our investigation, hoping to facilitate additional analyses, replication efforts, and discussion. The dataset, for example, allows researchers to extend our analysis to other racial groups, as well as rural versus urban counties.
\end{itemize}

\section{Observational design of our study}

At a high level, our goal is to determine whether the racial composition of a US county is associated with higher or lower visibility on Twitter's Home Timeline. This requires demographic information at the county level as well as a measure of visibility at the Twitter user level. Both are detailed below.

\subsection{Definition of county and associated demographics}

Our unit of analysis is the county or ``county equivalent'', such as Louisiana parishes. For simplicity we are going to use the word county to refer to both. Counties that are part of the study are from the 2018 American Community Survey (ACS) of the US Census Bureau~\cite{acs2018}. For the demographic data, we use racial and ethnic data from the same source~\cite{acs2018}, thus relying on the 2018 ACS categorization of race. Racial and ethnic data in the Census is self-identified, where participants can choose to mark one or more of the following categories: White, Black or African-American, American Indian or Alaska Native, Asian, Native Hawaiian or Other Pacific Islander. There is a separate question for ethnicity, which asks whether or not participants are of Hispanic, Latino, or Spanish origin or not.\footnote{We refer to the following features from the Census using the following labels: \texttt{DP05\_0037E}: white, \texttt{DP05\_0038E}: Black, \texttt{DP05\_0077E}: White alone, not Hispanic or Latino,
  \texttt{DP05\_0078E}: Black alone, not Hispanic or Latino. \\
  See \url{https://api.census.gov/data/2018/acs/acs5/profile/variables.html} for complete list of features in ACS 5 2018. For more information on the conceptualization of race on the Census, see \url{https://www.census.gov/topics/population/race/about.html}.}
Hispanic, Latino, or Spanish people may be of any race. For the limitations of this categorization, see Section~\ref{race}. 

\subsection{Time and scope of the data collection}

Our study includes all Twitter users who posted at last one Tweet between September 1, 2021 and September 9, 2021, that we were able to associate a location with in the United States. We assign to each user a unique county, which is the one from where they logged in the most frequently between August 10, 2021 and August 24, 2021. See Section~\ref{sec:location} for more details on location and its limitations. This study includes on the order of tens of millions of unique users.

\subsection{Impressions and amplification}

Throughout the paper, we use the concept of an \emph{impression}. When a Tweet is shown to any user in the Home Timeline, this counts as one \emph{impression} for the author of the Tweet\footnote{Specifically a Tweet is considered impressed by user A when 50\% of the UI element containing the Tweet is continuously visible on the user’s device for 0.5s.}. We consider \emph{unique} impressions, meaning that multiple impressions by the same reader count as one. We restrict impressions to readers who see Twitter's default algorithmically personalized timeline. Twitter users may opt out of personalization and a small fraction of users are randomly excluded from algorithmic personalization, see~\cite{doi:10.1073/pnas.2025334119} for more details on this aspect.

The number of impressions associated with a user are generally a function of how much content the user authors and how many followers they have. We make the simplifying assumption that the number of impressions scales bilinearly in the number of followers a user has and the number of Tweets the user authors. This assumption motivates the notion of a \emph{normalized impression}.

\begin{definition}
Given a user $u_i$ we define their \emph{normalized impressions} count with respect to a fixed time window as the total number of unique impressions received divided by the product of the total number of followers and the number of Tweets produced during the time window, i.e.,
\begin{equation}
\begin{aligned} 
& \mathrm{normalized\_impressions} 
= \\
& \frac{\mathrm{total\_unique\_impressions}}{(1 + \mathrm{total\_tweets\_produced}) \times (1+ \mathrm{num\_followers})}\,.
\end{aligned}
\end{equation}
Here, $\mathrm{total\_unique\_impressions}$ is the total number of unique impressions received only via the algorithmically-ranked timeline, \\
$\mathrm{ total\_tweets\_produced}$ is the total number of Tweets produced in the time frame, and $\mathrm{num\_followers}$ is the number of followers at the beginning of the experiment.\footnote{While the number of followers varies over time, on average it did not change much for the data collection period of roughly one week.}
\end{definition}

Normalized impressions are what allows us to define what an amplified user is.

\begin{definition}
We call a user \emph{amplified} if they have more normalized impressions than the US median user.
\end{definition}

Note that by definition of median, there is an equal number of amplified and non-amplified users. As a robustness check, we considered quantiles other than the median in the definition.

\subsection{Analysis 1: Effect of racial composition on amplification}
\label{sec:analysis1}

For each of the two racial groups we focus on (\emph{Black or African American} and \emph{White}), we fit one least squares linear regression model $Y=\alpha + \beta X,$ where $X$ is the fraction of the county's population in the given racial group and $Y$ is the share of amplified users in the county. Both variables range in the interval~$[0, 1].$ The unit of analysis is the county, meaning that each data point corresponds to one county.

We use the regression coefficient~$\beta$ as an observational measure of bias: A positive coefficient indicates that fractional size of the group within a county is associated with a higher share of amplified users. A negative coefficient indicates the opposite. To take into account the fact that different counties have different number of users (e.g. metropolitan areas tend to have more Twitter users), we perform two kinds of regression: an unweighted one (each county has equal weight) and weighted one, where the counties are weighted by the share of Twitter users (not the total population). We report the $\beta$-coefficient for each of the four regressions with $95\%$ confidence intervals, as well as the $R^2$-values.

In a two variable linear causal model $X\rightarrow Y$ with no unobserved confounders, the coefficient $\beta$ can be interpreted as the causal effect of a county's population in a given racial group on amplification. At the county level, a possible confounder could be a property of the county that influences both the county's racial composition and its amplification on Twitter. A plausible confounder that we did not control for is the county's urban-rural status, a classification also provided by the US Census Bureau.

\subsection{Analysis 2: Distribution of amplified users by county}
\label{sec:analysis2}

For our second analysis, we divide the counties into two separate sets, namely counties above and below the median of each racial group we consider. We then consider the histograms of the fraction of amplified users for each of the two cohorts. The difference between the two histograms is another indicator that racial composition of a county is associated with amplification. We report the mean and standard deviation for each distribution, as well as the total variation distance between the two distributions corresponding to the county split.

\subsection{Limitations}

Our study design has several inherent limitations and some technical limitations specific to the Twitter context.

\subsubsection{Location}
\label{sec:location}
The county of a user is constructed from available location data. The location of a user has intrinsic variation over time. Some users tag their Tweet with a location, which gives more confidence about their location at the time of writing the Tweet. There is lower precision when trying to infer the location of a user via IP address\footnote{More information on how Twitter uses location data can be found on \url{https://twitter.com/en/privacy}}.

As stated above, we are assigning to a user the county they log-in the most from, while the Census data measures the location where people live in. There could be discrepancies between the two: it could be possible that people who live in places with poor internet access will tend to use Twitter more in location with better internet access (e.g. the library, a work location etc), which might be located in a different county than the one where they reside. 

A small fraction of users that we aren't able to locate are dropped from the study.

\subsubsection{Data Loss}\label{data_loss}

The 2018 ACS Census data has 3219 counties listed that are part of our study. After assigning users to counties, we notice that 120 counties had no users assigned to them. Note that this need not imply that there are no users in those counties, just that we weren't able to associate users to them. In some cases this seems expected, as some of those are inhabited by a few hundred people and it's possible that no one used the platform from any of those. In other cases the size or population alone does not justify the lack of user data, and it's possible that data loss is happening at a step in the data pipeline. %The list of counties with missing users can be found in Table~\ref{table:missing} in Appendix~\ref{appendix:missing}.

\subsubsection{Level of Granularity}

Our analysis focuses on exposure disparity at the county level. From this data, we are not able to verify that our observations also hold at the user level. To illustrate the point, consider a county with a larger than average Black of African American population that has a large fraction of amplified users. Given the lack of demographic data on individual users, we don't know the race of amplified users \emph{within} the county. It is possible, in principle, that all amplified users within the county are not Black or African American. Testing this hypothesis would require racial information about individuals.

\subsubsection{Disparities between Census and Twitter's population}
When aggregating the users' population and comparing it with the Census data, we are implicitly assuming that the Census data is representative for Twitter's users. Research suggests that US-based Twitter's users are more young and liberal than the general population \cite{Hughes2019SizingUT}. However, we do not adjust for how the racial demographics of the general population in a county differ from those who use Twitter within the county.

\subsubsection{Definition of race} \label{race}

 Following the recommendations in recent work~\cite{hanna2020towards, scheuerman2020we}, we give a brief description of the sociohistorical context of the racial and ethnic data of the Census to elucidate the limitations of Census data. With strong influence from eugenicists and those looking to quantitatively validate the use chattel slavery, racial categories on the Census were developed in relation to Whiteness \cite{strmic2018race,hanna2020towards}. The taxonomy used by the Census is closely tied with the allocation of resources, but has also has been used to manage and surveil different populations, and thus the categories have been politically contested throughout its history \cite{strmic2018race,hanna2020towards}. For example, in 1930, “Mexican” was added to the Census in part due to increasing anti-Mexican sentiment during the great depression, but the Mexican government lobbied for it to be removed and for Mexicans to be classified as “White” in order to assert Mexicans should not be subject to Jim Crow laws in the US \cite{strmic2018race}. However, after civil rights, minority groups had more interest in being counted in order to fight discrimination and get access to resources . In 1980, the Census added a specific question that asked first about “Hispanic, Latino, or Spanish origin” followed by a separate race question. Proponents argue two separate questions capture two separate axes of oppression for race and ethnicity and better capture the heterogeneity of the Latino community. However, others argue combining the two questions could lead to a more reflective count of Latinos. More recently, because Arab Americans have been undercounted and disadvantaged in terms of acquiring resources that could help them, some have called for the addition of a MENA category (Middle Eastern and North African). However, given the post 9/11 political climate, others fear it could be used to increase surveillance and policing of this community \cite{strmic2018race}. 
These examples illustrate that the Census does not reflect natural or inherent categories, but rather constitutes and reinscribes socially constructed categories \cite{hanna2020towards,fields2014racecraft,benthall2019racial}. The use of racial/ethnic data here is not meant to reify racial/ethnic categories as natural, but rather study the impact of algorithmic harms on historical marginalized groups \cite{hanna2020towards,noble2018algorithms}.  
% Additionally, its worth noting that operationalizing race along a Black/white binary is an extremely US-centric conceptualization of race, and in other countries analyzing different attributes would be more appropriate. irst of all how race is constructed and perceived is a deeply social construct, that is very depend on the country. Understanding on the historical and current power structure are fundamental to make sense of the data.

\subsubsection{Amplification from other sources}
We are only considering Tweets that are scored (i.e. position-manipulated) by the Home Timeline ranking systems\footnote{For more information about the Home Timeline please refer to \url{https://help.twitter.com/en/using-twitter/twitter-timeline}}. If certain users get more exposure from other sources (for example ads or ``Who to Follow'' recommendations), such exposure will not be captured in the analysis. Similarly, if an author has a large number of followers who opted into the reverse-chronologically sorted timelines, their normalized impressions may be lower than expected. When computing the ratio, the numerator is not going to include the impressions from those users, while the denominator will stay the same.

\section{Analysis and observations} \label{results}

In this section, we present the findings of the two biases analyses we conducted. The first is the regression analysis described in Section~\ref{sec:analysis1}. The second is the distribution comparison outlined in Section~\ref{sec:analysis2}.

\subsection{Linear relationship between racial composition on amplification}

In our first analysis, we show the linear fit between the percentage of the population in a given racial group and the share of amplified users in a county. 

Table~\ref{tab:regression_coeff} shows the results of the regression analysis. We run the regression using the percentage of each racial group to predict the share of amplified users. We run both the weighted and unweighted regressions, where the weights are the percentage of Twitter users in that county (as a share of the US total). We do report the regression coefficients, associated 95\% confidence interval, and the $R^2$ value (i.e. the share of the variance explained by the linear model). 

\begin{table*}[h]
\begin{tabular}{llrr}
Independent variable &  Statistic  &  Weighted &  Unweighted \\
\midrule
Percentage of Black population& Coefficient &     
  0.0524 &    0.0593 \\
& 95\% CI &  [0.0322, 0.0726] &    [0.0358, 0.0827]\\
& $R^2$ &  0.0083 &    0.0079 \\
%  0.052411 &    0.059258 \\
%& 0.025 CI &  0.032191 &    0.035816 \\
%& 0.975 CI &  0.072631 &    0.082700 \\
%& $R^2$ &  0.008271 &    0.007870 \\
Percentage of White population & Coefficient 
& -0.1451 &   -0.0660 \\
& 95\% CI & [-0.1609, -0.1294] &   [-0.0862, -0.0458] \\
& $R^2$ &  0.0954 &    0.0131 \\
%-0.145149 &   -0.065966 \\
%& 0.025 CI & -0.160900 &   -0.086150 \\
%& 0.975  CI & -0.129397 &   -0.045782 \\
%& $R^2$ &  0.095357 &    0.013086 \\
\bottomrule
\end{tabular}

\caption{Regression coefficients, confidence intervals, and $R^2$ values}
\label{tab:regression_coeff}
\end{table*}

\begin{figure}[h]
\begin{subfigure}{0.44\textwidth}
\includegraphics[width=0.9\linewidth]{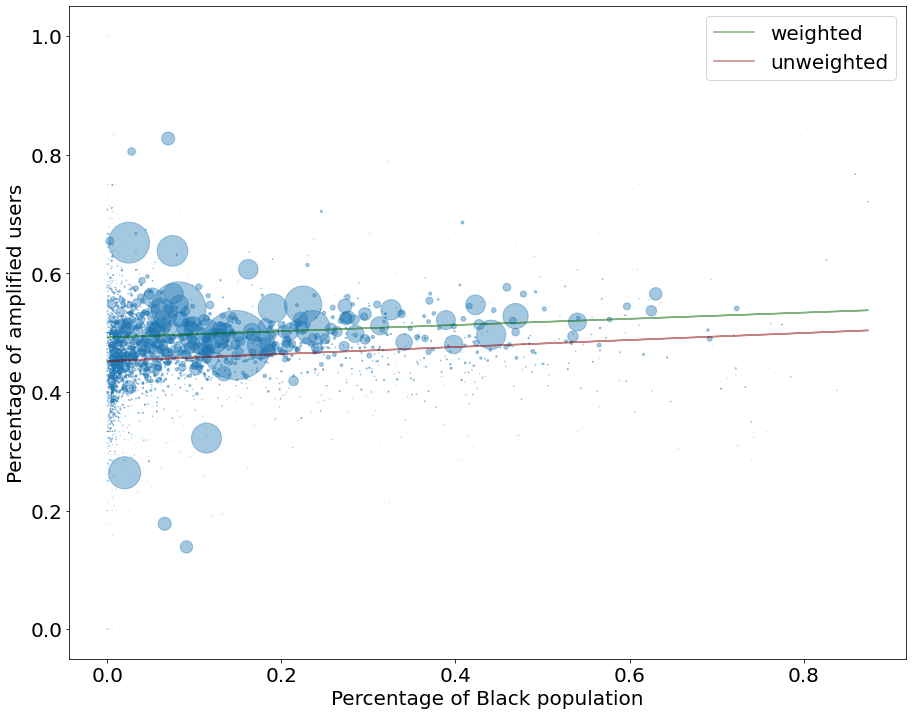} 
\caption{Black population}
\label{fig:scatterblack}
\end{subfigure}
\begin{subfigure}{0.44\textwidth}
\includegraphics[width=0.9\linewidth]{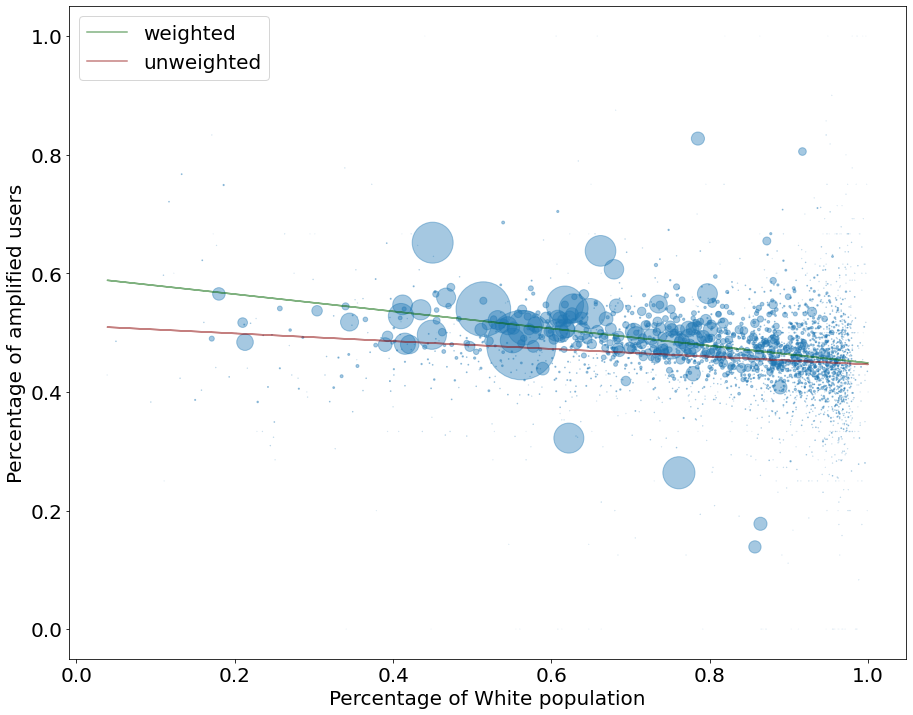}
\caption{White population}
\label{fig:scatterwhite}
\end{subfigure}
\caption{Distribution of share of amplified users between the top and bottom half for each race group.}
\label{fig:racehisto}
\end{figure}

Figure~\ref{fig:scatterblack} and Figure~\ref{fig:scatterwhite} visualize the data via scatter plots, where each circle corresponds to county scaled by the size of its population. The y-axis shows the percentage of the population in a racial group, while the x-axis presents the fraction of amplified users.

\subsection{Distribution of amplified users by county and racial group}

Per each racial group, we split the counties on the US median value, obtaining two groups of almost equal size, see below and Section~\ref{data_loss} for more details. We then measure the distance between the two resulting distributions using the total variation distance. We report the results in Table~\ref{fig:hist_diff}.

\begin{figure}[h]
\begin{subfigure}{0.44\textwidth}
\includegraphics[width=0.9\linewidth]{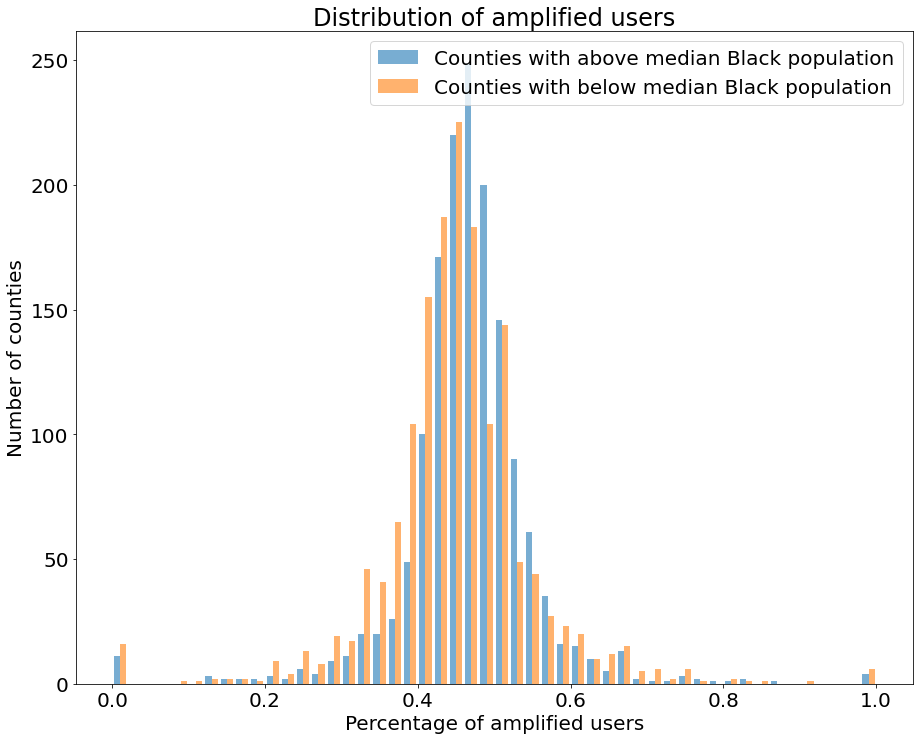} 
\caption{Black population}
%\label{fig:subim1}
\end{subfigure}
\begin{subfigure}{0.44\textwidth}
\includegraphics[width=0.9\linewidth]{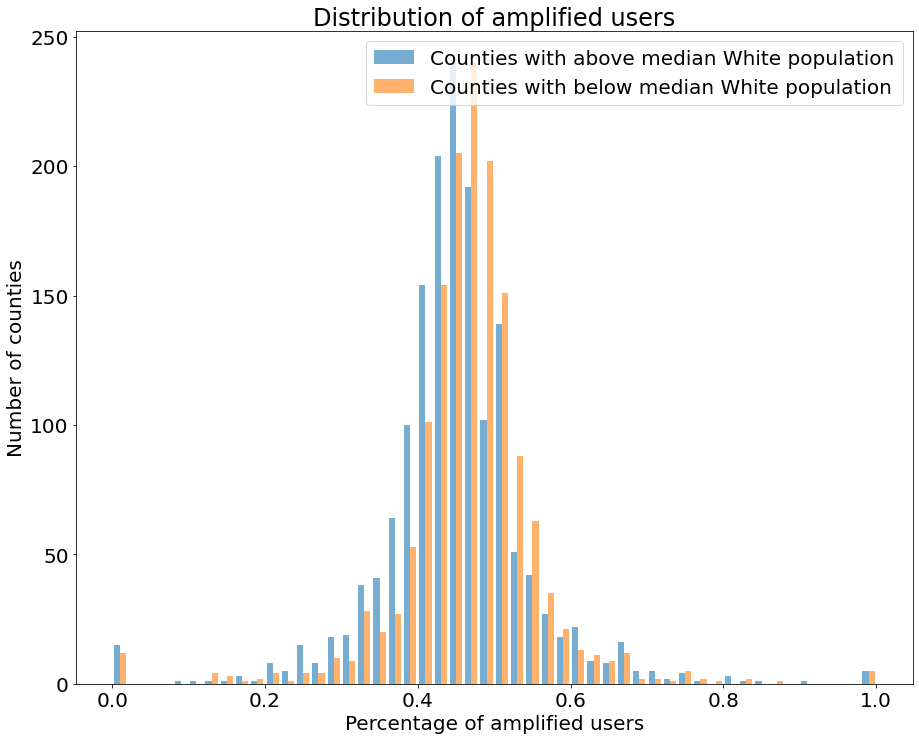}
\caption{White population}
%\label{fig:subim2}
\end{subfigure}
\caption{Distribution of share of amplified users between the top and bottom half for each race group.}
\label{fig:racehisto}
\end{figure}

\begin{table*}[h]
\begin{tabular}{lrr}

{} &  Black &  White \\
\midrule
Total variation &        0.172684 &         0.18839 \\
%\bottomrule
\end{tabular}
\caption{Measure of the difference between the top and bottom quantile for each race dimension.\label{fig:hist_diff}}
\end{table*}
\begin{table*}[h]
\begin{tabular}{lrrrr}

{} & \multicolumn{2}{c}{Black} & \multicolumn{2}{c}{White}   \\
{} &          Above Median&        Below Median &          Above Median &        Below Median \\
\midrule
Mean     &      30.3800 &    31.6000 &      31.8200 &    30.1600                                    \\
Variance &    3659.1956 &  3028.6000 &    3276.5876 &  3415.0944                                  \\
Std err  &       8.6416 &     7.8618 &       8.1773 &     8.3484                                     \\
\bottomrule
\end{tabular}
\caption{Summary statistics of the above histograms.\label{fig:hist_summ}}
\end{table*}

There is a caveat to add at this point. While, by definition, there is an equal number of counties in in below and above the median, this does not translate in an equal number of counties when we plot the above histograms (and compute their distances). The reason is that there are 120 counties missing in the final dataset as we could not associate any users to them for reasons described in Section \ref{data_loss}. Out of those, 18 are in the top quantile for the White population and 102 in the bottom one. For the Black population, 90 are in the top quantile and 30 in the bottom one.  

\subsection{Dataset description}

As part of this report, we are releasing the aggregated data used for our analysis. The dataset includes the 3099 counties that we have data for, with one county per row. The columns correspond to features of the county, most provided by the Census data. Out of the 121 columns, only two (\emph{twitter\_amplified\_users\_percent} and \emph{twitter\_users\_share}) are based on Twitter data. We are leaving the Census data in the released dataset for ease of use. The meaning of the two new columns is as follows:
\begin{itemize}
    \item \emph{twitter\_amplified\_users\_percent}: share of amplified users in each county. The number is between 0 and 1. For example a value of .65 means that 65\% of the users in that county are amplified, i.e. have above median normalized impression. 
    \item \emph{twitter\_users\_share}: share of Twitter's US-based users that are assigned to the county. The number is between 0 and 1. For example a value of 0.014 would mean that 1.4\% of all US-based users have been assigned to the county.
\end{itemize}

\section{Discussion}

Our analysis is based on counting impressions of Tweets in different counties. As such it gives a coarse representation of visibility on Twitter's Home Timeline. 
We see our analysis as providing a piece of quantitative evidence in a growing ecosystem of algorithmic audits of online platforms. Many questions remain.

The analysis is based on roughly a week's worth of Tweets. It does not begin to answer how possibly small differences in visibility compound over time as expressed in differential growth in the number of followers or diverging usage patterns.

Implicit in our analysis is the assumptions that an unbiased world is one in which each county has the same fraction of amplified users. But deviations from this baseline state of equality may have different causes and interpretations. Counties varies in many ways. Some urban areas might inherently feature more events that people Tweet about. Different groups also use Twitter differently, as illustrated in a manner relevant to our work by Brock's account of African American culture on Twitter~\cite{brock12from, brock2020distributed}. 

Visibility itself is not necessarily a normative goal. Some groups of users may use Twitter for conversations that are not meant to attract a wider audience. Others actively seek a broad audience for their message on Twitter.

When users see Tweets on the Home timeline, they come from a variety of sources: people they directly follow, Topics they follow, Tweets from people they do not follow imputed by personalization algorithms (e.g., based on previous likes), ads, and other kinds of content. It would be interesting to understand what impact each type of suggestion has on the overall amplification. However, it is less clear how to normalize impressions for these different content types. As an example, not all the Tweets liked by someone a user follows will be considered for a suggestion on the user's Home Timeline.

%It's also very unclear who should be considered a candidate for each content type. As a very extreme case, let's consider someone who only Tweets once, with a very modest following. We would not expect their Tweets to be recommended to users outside their network. One could then decide to only consider users who received impression via the specific impression type, but this analysis would also have serious limitations. Let's assume that the suggestion type excludes everyone from a specific group; this methodology will show no bias at all, as only successful users would be included. A different methodology would be required to better understand how each component plays a role.

The methodology we follow does not necessarily generalize to other regions. Because the US census offers a conceptualization of race that is unique to the historical and current power structure of the United States, if this approach of using location and census data was to be extended to other countries, analyzing different attributes would likely be more appropriate. Additionally, while using location based data might make sense for the United States---where people tend to live close to people with similar traits---it might not generalize well to other countries. In countries without the same history of segregation or where housing is more heterogeneous with respect to sensitive demographic traits, residential location data may not be as useful as a proxy for bias analysis~\cite{sambasivan2021re}.

Not least our report gives a glimpse at the challenges of measuring racial bias on online platforms, especially in the absence of individual-level racial information. We highlight an unaddressed tension between data privacy and algorithmic bias. In other words—the best way to analyze for bias based on a characteristic is to have that information, but the best way to ensure appropriate use is to never collect that data at all.

\bibliographystyle{ACM-Reference-Format}
\bibliography{biblio}


\begin{thebibliography}{24}

%%% the Web address.

\ifx \showCODEN    \undefined \def \showCODEN     #1{\unskip}     \fi
\ifx \showDOI      \undefined \def \showDOI       #1{#1}\fi
\ifx \showISBNx    \undefined \def \showISBNx     #1{\unskip}     \fi
\ifx \showISBNxiii \undefined \def \showISBNxiii  #1{\unskip}     \fi
\ifx \showISSN     \undefined \def \showISSN      #1{\unskip}     \fi
\ifx \showLCCN     \undefined \def \showLCCN      #1{\unskip}     \fi
\ifx \shownote     \undefined \def \shownote      #1{#1}          \fi
\ifx \showarticletitle \undefined \def \showarticletitle #1{#1}   \fi
\ifx \showURL      \undefined \def \showURL       {\relax}        \fi
% invisible to TeX
\providecommand\bibfield[2]{#2}
\providecommand\bibinfo[2]{#2}
\providecommand\natexlab[1]{#1}
\providecommand\showeprint[2][]{arXiv:#2}

\bibitem[Andrus et~al\mbox{.}(2021)]%
        {andrus2021we}
\bibfield{author}{\bibinfo{person}{McKane Andrus}, \bibinfo{person}{Elena
  Spitzer}, \bibinfo{person}{Jeffrey Brown}, {and} \bibinfo{person}{Alice
  Xiang}.} \bibinfo{year}{2021}\natexlab{}.
\newblock \showarticletitle{What We Can't Measure, We Can't Understand:
  Challenges to Demographic Data Procurement in the Pursuit of Fairness}. In
  \bibinfo{booktitle}{\emph{Proceedings of the 2021 ACM Conference on Fairness,
  Accountability, and Transparency}}. \bibinfo{pages}{249--260}.
\newblock


\bibitem[Benthall and Haynes(2019)]%
        {benthall2019racial}
\bibfield{author}{\bibinfo{person}{Sebastian Benthall} {and}
  \bibinfo{person}{Bruce~D Haynes}.} \bibinfo{year}{2019}\natexlab{}.
\newblock \showarticletitle{Racial categories in machine learning}. In
  \bibinfo{booktitle}{\emph{Proceedings of the conference on fairness,
  accountability, and transparency}}. \bibinfo{pages}{289--298}.
\newblock


\bibitem[Brock(2012)]%
        {brock12from}
\bibfield{author}{\bibinfo{person}{André Brock}.}
  \bibinfo{year}{2012}\natexlab{}.
\newblock \showarticletitle{From the Blackhand Side: Twitter as a Cultural
  Conversation}.
\newblock \bibinfo{journal}{\emph{Journal of Broadcasting \& Electronic Media}}
  \bibinfo{volume}{56}, \bibinfo{number}{4} (\bibinfo{year}{2012}),
  \bibinfo{pages}{529--549}.
\newblock


\bibitem[Brock(2020)]%
        {brock2020distributed}
\bibfield{author}{\bibinfo{person}{Andr{\'e} Brock}.}
  \bibinfo{year}{2020}\natexlab{}.
\newblock \showarticletitle{Distributed blackness}.
\newblock In \bibinfo{booktitle}{\emph{Distributed Blackness}}.
  \bibinfo{publisher}{New York University Press}.
\newblock


\bibitem[Bureau(2018)]%
        {acs2018}
\bibfield{author}{\bibinfo{person}{U.S.~Census Bureau}.}
  \bibinfo{year}{2018}\natexlab{}.
\newblock \bibinfo{title}{American Community Survey 5-Year Data}.
\newblock
\newblock
\urldef\tempurl%
\url{https://www.census.gov/data/developers/data-sets/acs-5year.html}
\showURL{%
\tempurl}


\bibitem[Chang et~al\mbox{.}(2021)]%
        {chang2021targeted}
\bibfield{author}{\bibinfo{person}{Ho-Chun~Herbert Chang},
  \bibinfo{person}{Matt Bui}, {and} \bibinfo{person}{Charlton McIlwain}.}
  \bibinfo{year}{2021}\natexlab{}.
\newblock \showarticletitle{Targeted Ads and/as Racial Discrimination:
  Exploring Trends in New York City Ads for College Scholarships}.
\newblock \bibinfo{journal}{\emph{arXiv preprint arXiv:2109.15294}}
  (\bibinfo{year}{2021}).
\newblock


\bibitem[Fields and Fields(2014)]%
        {fields2014racecraft}
\bibfield{author}{\bibinfo{person}{Karen~E Fields} {and}
  \bibinfo{person}{Barbara~J Fields}.} \bibinfo{year}{2014}\natexlab{}.
\newblock \bibinfo{booktitle}{\emph{Racecraft: The soul of inequality in
  American life}}.
\newblock \bibinfo{publisher}{Verso Books}.
\newblock


\bibitem[Freedman(1999)]%
        {freedman1999ecological}
\bibfield{author}{\bibinfo{person}{David~A Freedman}.}
  \bibinfo{year}{1999}\natexlab{}.
\newblock \showarticletitle{Ecological inference and the ecological fallacy}.
\newblock \bibinfo{journal}{\emph{International Encyclopedia of the social \&
  Behavioral sciences}} \bibinfo{volume}{6}, \bibinfo{number}{4027-4030}
  (\bibinfo{year}{1999}), \bibinfo{pages}{1--7}.
\newblock


\bibitem[Geronimus et~al\mbox{.}(1996)]%
        {geronimus1996validity}
\bibfield{author}{\bibinfo{person}{Arline~T Geronimus}, \bibinfo{person}{John
  Bound}, {and} \bibinfo{person}{Lisa~J Neidert}.}
  \bibinfo{year}{1996}\natexlab{}.
\newblock \showarticletitle{On the validity of using census geocode
  characteristics to proxy individual socioeconomic characteristics}.
\newblock \bibinfo{journal}{\emph{J. Amer. Statist. Assoc.}}
  \bibinfo{volume}{91}, \bibinfo{number}{434} (\bibinfo{year}{1996}),
  \bibinfo{pages}{529--537}.
\newblock


\bibitem[Gornick et~al\mbox{.}(1996)]%
        {gornick1996effects}
\bibfield{author}{\bibinfo{person}{Marian~E Gornick}, \bibinfo{person}{Paul~W
  Eggers}, \bibinfo{person}{Thomas~W Reilly}, \bibinfo{person}{Renee~M
  Mentnech}, \bibinfo{person}{Leslye~K Fitterman}, \bibinfo{person}{Lawrence~E
  Kucken}, {and} \bibinfo{person}{Bruce~C Vladeck}.}
  \bibinfo{year}{1996}\natexlab{}.
\newblock \showarticletitle{Effects of race and income on mortality and use of
  services among Medicare beneficiaries}.
\newblock \bibinfo{journal}{\emph{New England journal of medicine}}
  \bibinfo{volume}{335}, \bibinfo{number}{11} (\bibinfo{year}{1996}),
  \bibinfo{pages}{791--799}.
\newblock


\bibitem[Hanna et~al\mbox{.}(2020)]%
        {hanna2020towards}
\bibfield{author}{\bibinfo{person}{Alex Hanna}, \bibinfo{person}{Emily Denton},
  \bibinfo{person}{Andrew Smart}, {and} \bibinfo{person}{Jamila Smith-Loud}.}
  \bibinfo{year}{2020}\natexlab{}.
\newblock \showarticletitle{Towards a critical race methodology in algorithmic
  fairness}. In \bibinfo{booktitle}{\emph{Proceedings of the 2020 conference on
  fairness, accountability, and transparency}}. \bibinfo{pages}{501--512}.
\newblock


\bibitem[Huszár et~al\mbox{.}(2022)]%
        {doi:10.1073/pnas.2025334119}
\bibfield{author}{\bibinfo{person}{Ferenc Huszár}, \bibinfo{person}{Sofia~Ira
  Ktena}, \bibinfo{person}{Conor O’Brien}, \bibinfo{person}{Luca Belli},
  \bibinfo{person}{Andrew Schlaikjer}, {and} \bibinfo{person}{Moritz Hardt}.}
  \bibinfo{year}{2022}\natexlab{}.
\newblock \showarticletitle{Algorithmic amplification of politics on Twitter}.
\newblock \bibinfo{journal}{\emph{Proceedings of the National Academy of
  Sciences}} \bibinfo{volume}{119}, \bibinfo{number}{1} (\bibinfo{year}{2022}),
  \bibinfo{pages}{e2025334119}.
\newblock
\urldef\tempurl%
\url{https://doi.org/10.1073/pnas.2025334119}
\showDOI{\tempurl}
\showeprint{https://www.pnas.org/doi/pdf/10.1073/pnas.2025334119}


\bibitem[Johnson et~al\mbox{.}(2017)]%
        {johnson2017effect}
\bibfield{author}{\bibinfo{person}{Isaac Johnson}, \bibinfo{person}{Connor
  McMahon}, \bibinfo{person}{Johannes Sch{\"o}ning}, {and}
  \bibinfo{person}{Brent Hecht}.} \bibinfo{year}{2017}\natexlab{}.
\newblock \showarticletitle{The effect of population and" structural" biases on
  social media-based algorithms: A case study in geolocation inference across
  the urban-rural spectrum}. In \bibinfo{booktitle}{\emph{Proceedings of the
  2017 CHI conference on Human Factors in Computing Systems}}.
  \bibinfo{pages}{1167--1178}.
\newblock


\bibitem[Kallus et~al\mbox{.}(2022)]%
        {kallus2022assessing}
\bibfield{author}{\bibinfo{person}{Nathan Kallus}, \bibinfo{person}{Xiaojie
  Mao}, {and} \bibinfo{person}{Angela Zhou}.} \bibinfo{year}{2022}\natexlab{}.
\newblock \showarticletitle{Assessing algorithmic fairness with unobserved
  protected class using data combination}.
\newblock \bibinfo{journal}{\emph{Management Science}} \bibinfo{volume}{68},
  \bibinfo{number}{3} (\bibinfo{year}{2022}), \bibinfo{pages}{1959--1981}.
\newblock


\bibitem[Leitner et~al\mbox{.}(2016)]%
        {leitner2016racial}
\bibfield{author}{\bibinfo{person}{Jordan~B Leitner}, \bibinfo{person}{Eric
  Hehman}, \bibinfo{person}{Ozlem Ayduk}, {and} \bibinfo{person}{Rodolfo
  Mendoza-Denton}.} \bibinfo{year}{2016}\natexlab{}.
\newblock \showarticletitle{Racial bias is associated with ingroup death rate
  for Blacks and Whites: Insights from Project Implicit}.
\newblock \bibinfo{journal}{\emph{Social Science \& Medicine}}
  \bibinfo{volume}{170} (\bibinfo{year}{2016}), \bibinfo{pages}{220--227}.
\newblock


\bibitem[Monnat et~al\mbox{.}(2019)]%
        {monnat2019using}
\bibfield{author}{\bibinfo{person}{Shannon~M Monnat}, \bibinfo{person}{David~J
  Peters}, \bibinfo{person}{Mark~T Berg}, {and} \bibinfo{person}{Andrew
  Hochstetler}.} \bibinfo{year}{2019}\natexlab{}.
\newblock \showarticletitle{Using census data to understand county-level
  differences in overall drug mortality and opioid-related mortality by opioid
  type}.
\newblock \bibinfo{journal}{\emph{American Journal of Public Health}}
  \bibinfo{volume}{109}, \bibinfo{number}{8} (\bibinfo{year}{2019}),
  \bibinfo{pages}{1084--1091}.
\newblock


\bibitem[Noble(2018)]%
        {noble2018algorithms}
\bibfield{author}{\bibinfo{person}{Safiya~Umoja Noble}.}
  \bibinfo{year}{2018}\natexlab{}.
\newblock \bibinfo{booktitle}{\emph{Algorithms of oppression}}.
\newblock \bibinfo{publisher}{New York University Press}.
\newblock


\bibitem[Riddle and Sinclair(2019)]%
        {riddle2019racial}
\bibfield{author}{\bibinfo{person}{Travis Riddle} {and} \bibinfo{person}{Stacey
  Sinclair}.} \bibinfo{year}{2019}\natexlab{}.
\newblock \showarticletitle{Racial disparities in school-based disciplinary
  actions are associated with county-level rates of racial bias}.
\newblock \bibinfo{journal}{\emph{Proceedings of the National Academy of
  Sciences}} \bibinfo{volume}{116}, \bibinfo{number}{17}
  (\bibinfo{year}{2019}), \bibinfo{pages}{8255--8260}.
\newblock


\bibitem[Ross(2015)]%
        {ross2015multi}
\bibfield{author}{\bibinfo{person}{Cody~T Ross}.}
  \bibinfo{year}{2015}\natexlab{}.
\newblock \showarticletitle{A multi-level Bayesian analysis of racial bias in
  police shootings at the county-level in the United States, 2011--2014}.
\newblock \bibinfo{journal}{\emph{PloS one}} \bibinfo{volume}{10},
  \bibinfo{number}{11} (\bibinfo{year}{2015}), \bibinfo{pages}{e0141854}.
\newblock


\bibitem[Sambasivan et~al\mbox{.}(2021)]%
        {sambasivan2021re}
\bibfield{author}{\bibinfo{person}{Nithya Sambasivan}, \bibinfo{person}{Erin
  Arnesen}, \bibinfo{person}{Ben Hutchinson}, \bibinfo{person}{Tulsee Doshi},
  {and} \bibinfo{person}{Vinodkumar Prabhakaran}.}
  \bibinfo{year}{2021}\natexlab{}.
\newblock \showarticletitle{Re-imagining algorithmic fairness in india and
  beyond}. In \bibinfo{booktitle}{\emph{Proceedings of the 2021 ACM Conference
  on Fairness, Accountability, and Transparency}}. \bibinfo{pages}{315--328}.
\newblock


\bibitem[Scheuerman et~al\mbox{.}(2020)]%
        {scheuerman2020we}
\bibfield{author}{\bibinfo{person}{Morgan~Klaus Scheuerman},
  \bibinfo{person}{Kandrea Wade}, \bibinfo{person}{Caitlin Lustig}, {and}
  \bibinfo{person}{Jed~R Brubaker}.} \bibinfo{year}{2020}\natexlab{}.
\newblock \showarticletitle{How we've taught algorithms to see identity:
  constructing race and gender in image databases for facial analysis}.
\newblock \bibinfo{journal}{\emph{Proceedings of the ACM on Human-computer
  Interaction}} \bibinfo{volume}{4}, \bibinfo{number}{CSCW1}
  (\bibinfo{year}{2020}), \bibinfo{pages}{1--35}.
\newblock


\bibitem[Soobader et~al\mbox{.}(2001)]%
        {soobader2001using}
\bibfield{author}{\bibinfo{person}{Mah-jabeen Soobader},
  \bibinfo{person}{Felicia~B LeClere}, \bibinfo{person}{Wilbur Hadden}, {and}
  \bibinfo{person}{Brooke Maury}.} \bibinfo{year}{2001}\natexlab{}.
\newblock \showarticletitle{Using aggregate geographic data to proxy individual
  socioeconomic status: does size matter?}
\newblock \bibinfo{journal}{\emph{American Journal of Public Health}}
  \bibinfo{volume}{91}, \bibinfo{number}{4} (\bibinfo{year}{2001}),
  \bibinfo{pages}{632}.
\newblock


\bibitem[Strmic-Pawl et~al\mbox{.}(2018)]%
        {strmic2018race}
\bibfield{author}{\bibinfo{person}{Hephzibah~V Strmic-Pawl},
  \bibinfo{person}{Brandon~A Jackson}, {and} \bibinfo{person}{Steve Garner}.}
  \bibinfo{year}{2018}\natexlab{}.
\newblock \showarticletitle{Race counts: racial and ethnic data on the US
  Census and the implications for tracking inequality}.
\newblock \bibinfo{journal}{\emph{Sociology of Race and Ethnicity}}
  \bibinfo{volume}{4}, \bibinfo{number}{1} (\bibinfo{year}{2018}),
  \bibinfo{pages}{1--13}.
\newblock


\bibitem[Wojcik and Hughes(2019)]%
        {Hughes2019SizingUT}
\bibfield{author}{\bibinfo{person}{Stefan Wojcik} {and} \bibinfo{person}{Adam
  Hughes}.} \bibinfo{year}{2019}\natexlab{}.
\newblock \showarticletitle{Sizing Up Twitter Users}.
\newblock


\end{thebibliography}
% yes, they are not in order
\appendix{} 
\section{Results for other racial groups} \label{other_results}
We are presenting the same results as Section \ref{results} but for  other racial groups, namely \texttt{DP05\_0078E} (Black alone, not Hispanic or Latino) and \texttt{DP05\_0077E} (white alone, not Hispanic or Latino). 
As before we are starting with the scatter plots and the regression coefficients. 
 
 \begin{figure}[h]
\begin{subfigure}{0.44\textwidth}
\includegraphics[width=0.9\linewidth]{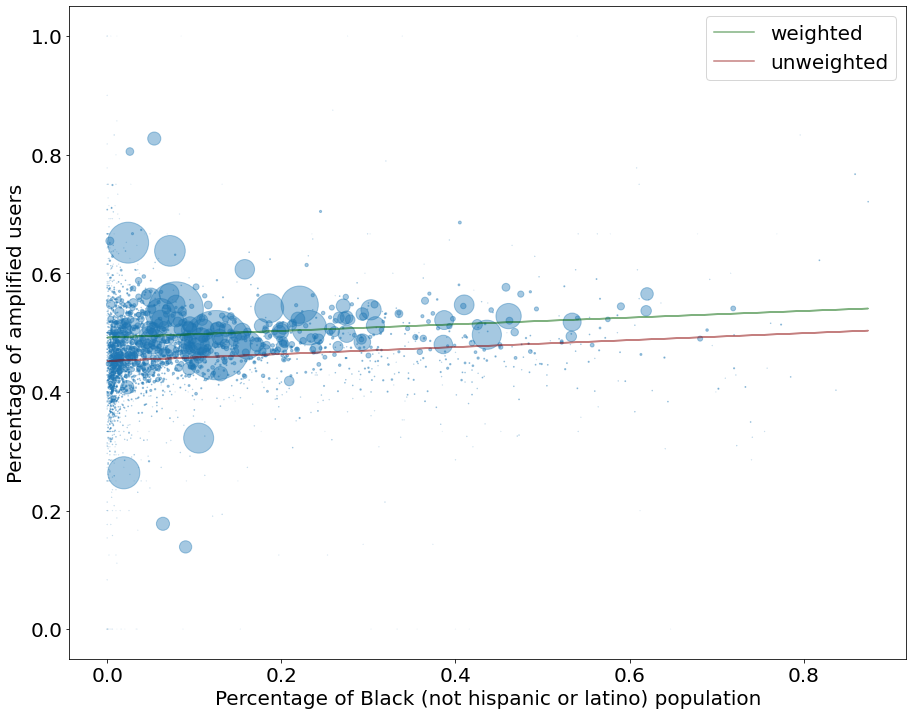} 
\caption{Black alone population}
%\label{fig:subim1}
\end{subfigure}
\begin{subfigure}{0.44\textwidth}
\includegraphics[width=0.9\linewidth]{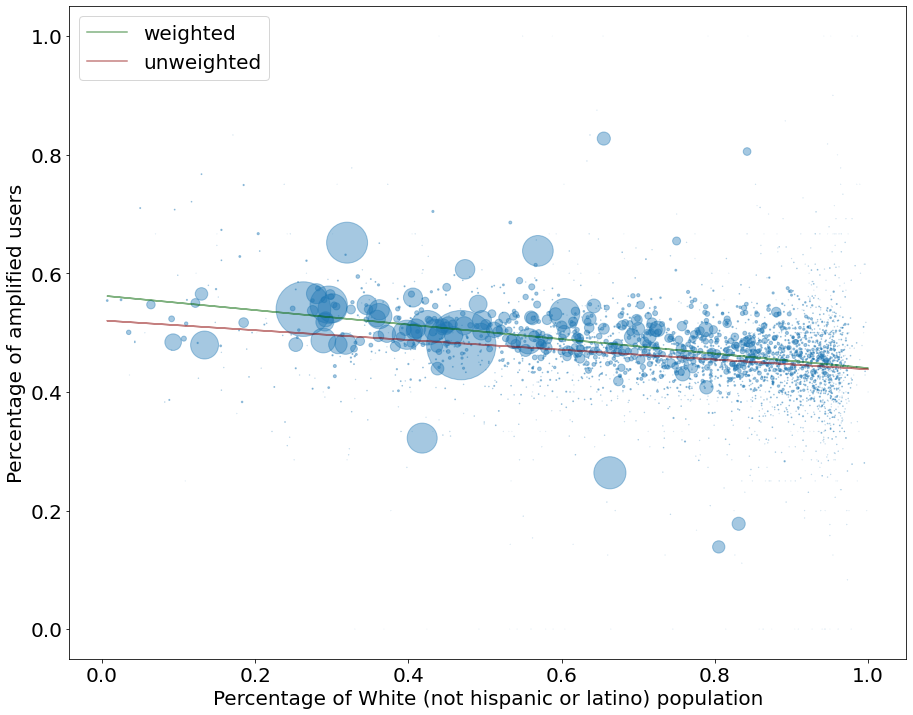}
\caption{White alone population}
%\label{fig:subim2}
\end{subfigure}
\caption{Distribution of share of amplified users between the top and bottom half for each race group.}
\label{fig:racehisto}
\end{figure}

\begin{table*}[h]
\begin{tabular}{llrr}
Independent variable &  Statistic  &  Weighted &  Unweighted \\
\midrule
Percentage of Black population, not Hispanic nor Latino & Coefficient &     
 0.0561 &    0.0587 \\
& 95\% CI &  [0.0356, 0.0765] &    [0.035, 0.0822]\\
& $R^2$ &  0.0092 &    0.0076 \\
%  0.052411 &    0.059258 \\
%& 0.025 CI &  0.032191 &    0.035816 \\
%& 0.975 CI &  0.072631 &    0.082700 \\
%& $R^2$ &  0.008271 &    0.007870 \\
Percentage of White population, not Hispanic nor Latino & Coefficient 
& -0.1222 &   -0.0830  \\
& 95\% CI & [-0.135,  -0.1102] &   [-0.0997, -0.0665] \\
& $R^2$ &   0.1115 &   0.0300  \\
%-0.145149 &   -0.065966 \\
%& 0.025 CI & -0.160900 &   -0.086150 \\
%& 0.975  CI & -0.129397 &   -0.045782 \\
%& $R^2$ &  0.095357 &    0.013086 \\
\bottomrule
\end{tabular}

\caption{Regression coefficients, confidence intervals, and $R^2$ values}
\label{tab:regression_coeff}
\end{table*}
 
Now we present the histograms and their relative statistics.
 \begin{figure}[h!]
\begin{subfigure}{0.44\textwidth}
\includegraphics[width=0.9\linewidth]{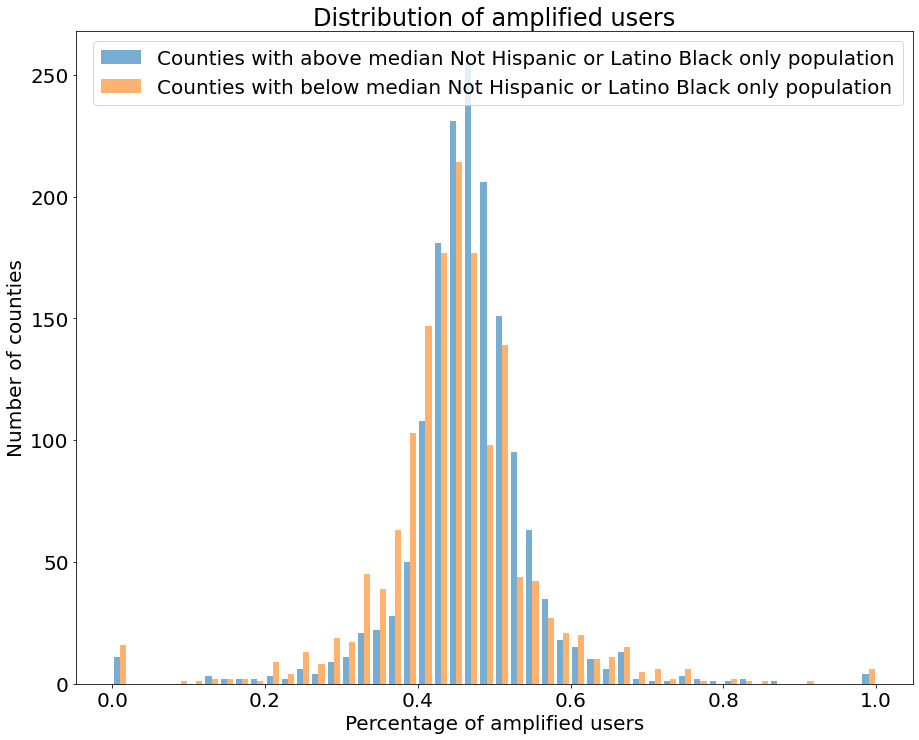} 
\caption{Black population}
%\label{fig:subim1}
\end{subfigure}
\begin{subfigure}{0.44\textwidth}
\includegraphics[width=0.9\linewidth]{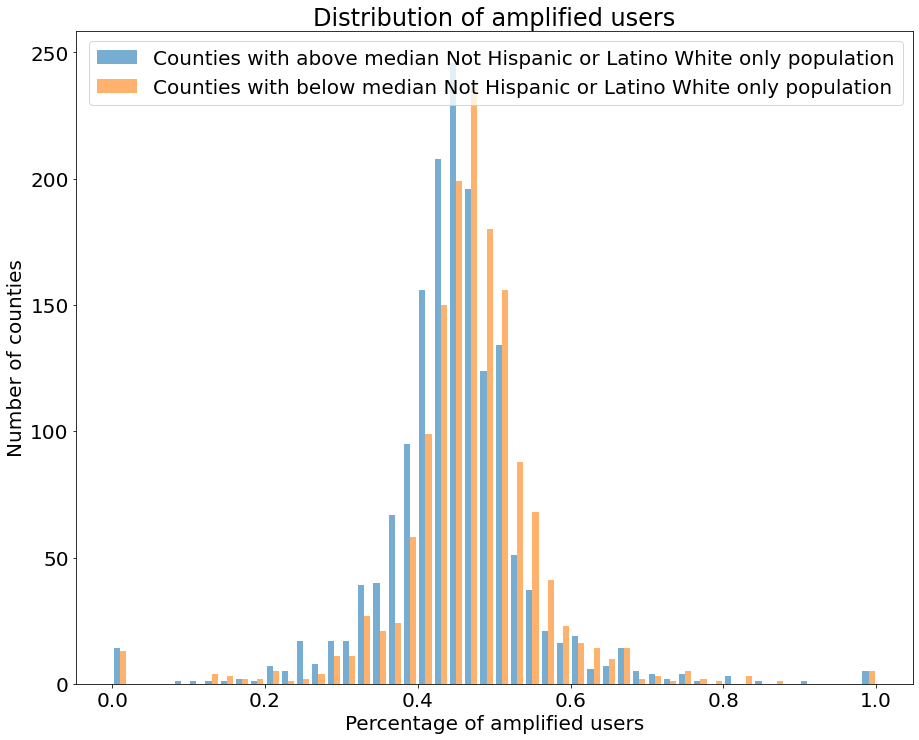}
\caption{White population}
%\label{fig:subim2}
\end{subfigure}
\caption{Distribution of share of amplified users between the top and bottom half for each race group.}
%\label{fig:racehisto}
\end{figure}

\begin{table*}[h]
\begin{tabular}[h]{lrr}

{} &    Black alone (no Hispanic or Latino) & White alone (no Hispanic or Latino)  \\
\midrule
Total Variation &    0.172727 &                    0.18306  \\
\bottomrule
\end{tabular}
\end{table*}

\begin{table*}[h]
\begin{tabular}{lrrrr}
 {} &
 \multicolumn{2}{c}{White alone (no Hispanic or Latino)} & \multicolumn{2}{c}{Black alone (no Hispanic or Latino)} \\
{} &          Above Median &        Below Median  &          Above Median  &        Below Median \\
\midrule
Mean     &                                       31.8800 &    30.1000 &                                  31.6200 &    30.3600 \\
Variance &    3445.9856 &  3200.7700 &                                3955.7956 &  2762.7504 \\
Std err  &        8.3860 &     8.0822 &                                   8.9850 &     7.5088 \\
\bottomrule
\end{tabular}
\caption{Summary statistics of the above histograms.\label{fig:hist_summ_apx}}
\end{table*}
% \begin{table}
% \begin{tabular}{llrr}
%                                       &     &  Weighted &  Unweighted \\
% \midrule
% Percentage of Black population, not Hispanic nor Latino  & Coefficient &  0.056058 &    0.058659 \\
%                                       & 0.025 CI&  0.035580 &    0.035105 \\
%                                       & 0.975 CI&  0.076536 &    0.082213 \\
%                                       & $R^2$ &  0.009217 &    0.007640 \\
% Percentage of white population, not Hispanic nor Latino  & Coefficient & -0.122377 &   -0.083198 \\
%                                       & 0.025 CI& -0.134549 &   -0.099874 \\
%                                       & 0.975 CI & -0.110205 &   -0.066522 \\
%                                       & $R^2$ &  0.111492 &    0.029973 \\

% \bottomrule
% \end{tabular}

% \caption{Regression coefficients.\label{regression_coeff}}
% \end{table}

\end{document}